\documentclass[11pt]{article}
\usepackage{amsmath,amssymb,amsthm,amsxtra,overpic}
\def\thefootnote{\fnsymbol{footnote}}

\begin{document}

\vspace{0.2cm}

\begin{center}
{\Large\bf Non-Hermitian Perturbations to the Fritzsch Textures of \\
Lepton and Quark Mass Matrices}
\end{center}

\vspace{0.1cm}

\begin{center}
{\bf Harald Fritzsch}$^\ddagger$, ~ {\bf Zhi-zhong Xing}$^\dagger$
\footnote{E-mail: xingzz@ihep.ac.cn}, ~ {\bf Ye-Ling Zhou}$^\dagger$ \\
$^\dagger${Institute of High Energy Physics, Chinese Academy of
Sciences, Beijing 100049, China} \\
$^\ddagger${Physik-Department, Universit$\rm\ddot{a}$t
M$\rm\ddot{u}$nchen, D-80333 Munich, Germany}
\end{center}

\vspace{1.5cm}

\begin{abstract}
We show that non-Hermitian and nearest-neighbor-interacting
perturbations to the Fritzsch textures of lepton and quark mass
matrices can make both of them fit current experimental data very
well. In particular, we obtain $\theta^{}_{23} \simeq 45^\circ$ for
the atmospheric neutrino mixing angle and predict $\theta^{}_{13}
\simeq 3^\circ$ to $6^\circ$ for the smallest neutrino mixing angle
when the perturbations in the lepton sector are at the $20\%$ level.
The same level of perturbations is required in the quark sector,
where the Jarlskog invariant of CP violation is about $3.7 \times
10^{-5}$. In comparison, the strength of leptonic CP violation is
possible to reach about $1.5 \times 10^{-2}$ in neutrino
oscillations.
\end{abstract}

\begin{flushleft}
\hspace{0.8cm} PACS number(s): 14.60.Pq, 13.10.+q, 25.30.Pt \\
\hspace{0.8cm} Keywords: quark and lepton masses, flavor mixing,
Fritzsch texture
\end{flushleft}

\def\thefootnote{\arabic{footnote}}
\setcounter{footnote}{0}

\newpage
\section{Introduction}

The flavor sector in the standard model (SM) of electroweak
interactions has been puzzling because it involves most of the free
parameters of the model itself. Although the values of six quark
masses and those of three angles and one CP-violating phase of the
$3\times 3$ Cabibbo-Kobayashi-Maskawa (CKM) quark mixing matrix $V$
\cite{CKM} are all known to a good degree of accuracy \cite{PDG}, it
remains very difficult to understand why they have the observed mass
spectrum and flavor mixing pattern. The situation in the lepton
sector is even worse: not only the absolute mass scale of three
neutrinos but also the smallest angle and CP-violating phases of the
$3\times 3$ Maki-Nakagawa-Sakata-Pontecorvo (MNSP) lepton mixing
matrix $U$ \cite{MNSP} are still unknown. To resolve the flavor
problem in the SM one has to gain an insight into the flavor
structures and possible flavor symmetries behind them. In spite of
many attempts in this direction, a successful (unique and
predictive) flavor theory has not been achieved. Most of the
present-day studies on the flavor structures of quarks and leptons
are more or less phenomenological \cite{Review}, and among them the
texture-zero approach \cite{Zero} has proved to be very useful to
establish some simple and testable relations between the mass ratios
of quarks or leptons and their corresponding flavor mixing angles.

In the three-family framework the Fritzsch texture of quark mass
matrices \cite{Fritzsch78}
\begin{eqnarray}
M^{({\rm F})}_\alpha = \begin{pmatrix} 0 & A^{}_\alpha & 0 \\
A^*_\alpha & 0 & B^{}_\alpha \\
0 & B^*_\alpha & C^{}_\alpha \end{pmatrix} \; ,
\end{eqnarray}
where $\alpha = {\rm u}$ (up) or $\rm d$ (down), has attracted a lot
of interest since it was proposed in 1978. It belongs to the more
generic nearest-neighbor-interaction (NNI) form of quark mass
matrices,
\begin{eqnarray}
M^{({\rm NNI})}_\alpha = \begin{pmatrix} 0 & A^{}_\alpha & 0 \\
A^\prime_\alpha & 0 & B^{}_\alpha \\
0 & B^\prime_\alpha & C^{}_\alpha \end{pmatrix} \; .
\end{eqnarray}
The NNI form can always be obtained from an arbitrary form of $M^{}_{\rm
u}$ and $M^{}_{\rm d}$ via a proper choice of the flavor basis in
the SM \cite{NNI}. So the Fritzsch texture is actually a NNI texture
with the additional assumption of the Hermiticity
conditions $A^\prime_\alpha = A^*_\alpha$ and $B^\prime_\alpha =
B^*_\alpha$. In view of the problem that $M^{({\rm F})}_{\rm u}$ and
$M^{({\rm F})}_{\rm d}$ cannot simultaneously give rise to a
sufficiently large value of the top-quark mass $m^{}_t$ and a
sufficiently small value of the CKM matrix element $|V^{}_{cb}|$,
one has to abandon either the NNI feature of quark mass matrices
\cite{Hamzaoui} or their Hermiticity \cite{Xing97}, either partly or
completely
\footnote{One may certainly abandon both the NNI and Hermiticity
conditions by following a different starting point of view (e.g.,
the triangular form of $M^{}_{\rm u}$ and $M^{}_{\rm d}$) to
investigate quark mass matrices and their consequences on flavor
mixing and CP violation \cite{Review}. But we do not focus on this
possibility in the present work.}.
It is always possible to numerically determine the departures of
realistic $M^{}_{\rm u}$ and $M^{}_{\rm d}$ from $M^{({\rm F})}_{\rm
u}$ and $M^{({\rm F})}_{\rm d}$ by using current experimental data.
If such departures are not very significant, they can be regarded as
small perturbations and then be treated in an analytical way so that
their effects on the CKM matrix elements will become more
transparent.

It also makes sense to consider non-Hermitian and
nearest-neighbor-interacting perturbations to the Fritzsch-type
lepton mass matrices $M^{({\rm F})}_l$ and $M^{({\rm F})}_\nu$.
Although the latter can fit current experimental data \cite{Xing02},
it is very difficult to obtain $\theta^{}_{23} \simeq 45^\circ$ or
equivalently the maximal or nearly maximal atmospheric neutrino
mixing. One possible way out is to introduce the seesaw mechanism to
the neutrino sector \cite{FTY}, in which the Dirac neutrino mass
matrix $M^{}_{\rm D}$ takes the Fritzsch texture as $M^{}_l$ does
but the heavy Majorana neutrino mass matrix $M^{}_{\rm R}$ is
(approximately) proportional to the identity matrix. Here we follow
a different way at the electroweak scale. We shall introduce
non-Hermitian perturbations to both $M^{({\rm F})}_l$ and $M^{({\rm
F})}_\nu$ so that the resultant charged-lepton and neutrino mass
matrices can agree with current neutrino oscillation data to a
better degree of accuracy
\footnote{In this treatment we have assumed massive neutrinos to be
the Dirac particles because the overall neutrino mass matrix
$M^{}_\nu$ is not symmetric. Of course, one may first apply the same
treatment to $M^{}_{\rm D}$ and then invoke the seesaw mechanism to
produce a Majorana mass matrix $M^{}_\nu$ for three light
neutrinos.}.
Such a parallel study of lepton and quark mass matrices of
approximate Fritzsch textures is useful to reveal the similarities
and differences between the lepton and quark sectors, and it should
also be helpful for building a unified flavor model of leptons and
quarks.

Let us point out that the present work is different from a recent one
done by Branco {\it et al} \cite{Branco} in the following three aspects.
(1) They have only considered quark mass matrices of the NNI form,
whereas we are discussing both lepton and quark mass matrices of the
same NNI form and giving a stronger emphasis to the lepton sector.
In particular, we are concerned about an interpretation of the
observed $\theta^{}_{23} \simeq 45^\circ$ and a prediction of nonzero $\theta^{}_{13}$ for the MNSP matrix based on the NNI texture of
lepton mass matrices at the electroweak scale. (2) The analytical
approximations made in our perturbative calculation are valid to a
better degree of accuracy and thus allow one to see the difference
between the contribution of fermion mass ratios and that of
perturbation parameters (which signify a departure of the NNI texture
from Hermiticity) to the flavor mixing matrix elements in a
clearer way. (3) Our numerical analysis is more comprehensive than the
one done in Ref. \cite{Branco}, and it shows that current experimental
data require the non-Hermitian effects to be at the $20\%$ level in
both lepton and quark sectors. This observation is expected to be useful
for model building, especially when leptons and quarks are
discussed in a unified flavor picture.

The remaining part of this paper is organized as follows. In section
2 we do a perturbative calculation to reveal the salient features of
non-Hermitian corrections to the Fritzsch textures of lepton and
quark mass matrices. Section 3 is devoted to a numerical
illustration of the constrained parameter space at a reasonable
level of perturbations in the lepton and quark sectors. A brief
summary, together with some further discussions, is given in section
4.

\section {A perturbative calculation}

Without loss of generality, the NNI mass matrix $M^{(\rm
NNI)}_\alpha$ (for $\alpha = {\rm u}, {\rm d}, l$ or $\nu$) in Eq.
(2) can always be decomposed into $M^{(\rm NNI)}_\alpha =
P^{}_\alpha \tilde{M}^{(\rm NNI)}_\alpha P^\prime_\alpha$, where
$P^{}_\alpha$ and $P^\prime_\alpha$ are two independent diagonal
phase matrices, and
\begin{eqnarray}
\tilde{M}^{({\rm NNI})}_\alpha = \begin{pmatrix} 0 & a^{}_\alpha & 0 \\
a^\prime_\alpha & 0 & b^{}_\alpha \\
0 & b^\prime_\alpha & c^{}_\alpha \end{pmatrix} \;
\end{eqnarray}
is real. After the bi-unitary transformation $O^\dagger_\alpha
\tilde{M}^{({\rm NNI})}_\alpha O^\prime_\alpha =
\widehat{M}^{}_\alpha \equiv {\rm Diag}\{\lambda^{\alpha}_1,
\lambda^{\alpha}_2, \lambda^{\alpha}_3 \}$ with $\lambda^{\alpha}_i$
(for $i=1,2,3$) being three mass eigenvalues, we can obtain the CKM
and MNSP matrices as follows:
\begin{eqnarray}
V \hspace{-0.2cm} &=& \hspace{-0.2cm} (P^{}_{\rm u} O^{}_{\rm
u})^\dagger (P^{}_{\rm d} O^{}_{\rm d})
= O^\dagger_{\rm u} P^{}_V O^{}_{\rm d} \; , \nonumber \\
U \hspace{-0.2cm} &=& \hspace{-0.2cm} (P^{}_l O^{}_l)^\dagger
(P^{}_\nu O^{}_\nu) = O^\dagger_l P^{}_U O^{}_\nu \; ,
\end{eqnarray}
where $P^{}_V \equiv P^\dagger_{\rm u} P^{}_{\rm d} = {\rm
Diag}\{e^{i\phi^{}_1}, e^{i\phi^{}_2}, 1\}$ and $P^{}_U \equiv
P^\dagger_l P^{}_\nu = {\rm Diag}\{e^{i\varphi^{}_1},
e^{i\varphi^{}_2}, 1\}$ are two diagonal phase matrices in a chosen
phase convention.

Following the same model-building strategy as specified in
Ref. \cite{Branco}, here we focus on the consequences of $M^{(\rm
NNI)}_\alpha$ on flavor mixing. We consider the real
mass matrix $\tilde{M}^{({\rm NNI})}_\alpha = \tilde{M}^{(\rm F)}_\alpha +
\tilde{M}^{(\epsilon)}_\alpha$, where
\begin{eqnarray}
\tilde{M}^{(\rm F)}_\alpha \hspace{-0.2cm} &=& \hspace{-0.2cm}
\begin{pmatrix} 0 & a^{}_\alpha & 0 \\
a^{}_\alpha & 0 & b^{}_\alpha \\
0 & b^{}_\alpha & c^{}_\alpha \end{pmatrix} \; , \nonumber \\
\tilde{M}^{(\epsilon)}_\alpha \hspace{-0.2cm} &=& \hspace{-0.2cm}
\begin{pmatrix} 0 & -a^{}_\alpha \epsilon^\alpha_a & 0 \\
+a^{}_\alpha \epsilon^\alpha_a & 0 & -b^{}_\alpha \epsilon^\alpha_b \\
0 & +b^{}_\alpha \epsilon^\alpha_b & 0 \end{pmatrix} \;
\end{eqnarray}
with $\epsilon^\alpha_a$ and $\epsilon^\alpha_b$ being dimensionless
real parameters describing small and asymmetric corrections to
$\tilde{M}^{(\rm F)}_\alpha$. Treating $\epsilon^\alpha_a$ and $\epsilon^\alpha_b$ as perturbation parameters will technically allow
us to perform an analytical diagonalization of
$\tilde{M}^{({\rm NNI})}_\alpha$.
For simplicity, we omit the flavor
index $\alpha$ in the subsequent discussions. It is easy to exactly
diagonalize $\tilde{M}^{(\rm NNI)} = \tilde{M}^{(\rm F)}$ in the
limit of $\epsilon^{}_a = \epsilon^{}_b =0$ \cite{Georgi}:
\begin{eqnarray}
a \hspace{-0.2cm} &=& \hspace{-0.2cm} \sqrt{\frac{\lambda_1^{}
\lambda_2^{} \lambda_3^{}} {\left(\lambda_1^{} - \lambda_2^{} +
\lambda_3^{}\right)}} \; ,
\nonumber\\
b \hspace{-0.2cm} &=& \hspace{-0.2cm} \sqrt{\frac{\left(\lambda_1^{}
- \lambda_2^{}\right) \left(\lambda_2^{} - \lambda_3^{}\right)
\left(\lambda_1^{} + \lambda_3^{}\right)}{\lambda_1^{} -
\lambda_2^{} + \lambda_3^{}}} \; ,
\nonumber\\
c \hspace{-0.2cm} &=& \hspace{-0.2cm}
\lambda_1^{} - \lambda_2^{} + \lambda_3^{} \; ;
\end{eqnarray}
and
\begin{eqnarray}
O^{(0)}_{11} \hspace{-0.2cm} &=& \hspace{-0.2cm}
\sqrt{\frac{\lambda_2^{} \lambda_3^{}(\lambda_3^{}-\lambda_2^{})
}{(\lambda_1^{}+\lambda_2^{})
(\lambda_1^{}-\lambda_2^{}+\lambda_3^{})
(\lambda_3^{}-\lambda_1^{})}} \; ,
\nonumber \\
O^{(0)}_{12} \hspace{-0.2cm} &=& \hspace{-0.2cm}
-\sqrt{\frac{\lambda_1^{} \lambda_3^{}
(\lambda_1^{}+\lambda_3^{})}{(\lambda_1^{}+\lambda_2^{})
(\lambda_1^{}-\lambda_2^{}+\lambda_3^{})
(\lambda_2^{}+\lambda_3^{})}} \; ,
\nonumber\\
O^{(0)}_{13} \hspace{-0.2cm} &=& \hspace{-0.2cm}
\sqrt{\frac{\lambda_1^{} \lambda_2^{}(\lambda_2^{}-\lambda_1^{})
}{(\lambda_3^{}-\lambda_1^{})
(\lambda_1^{}-\lambda_2^{}+\lambda_3^{})
(\lambda_2^{}+\lambda_3^{})}} \; ,
\nonumber\\
O^{(0)}_{21} \hspace{-0.2cm} &=& \hspace{-0.2cm}
\sqrt{\frac{\lambda_1^{}
(\lambda_3^{}-\lambda_2^{})}{(\lambda_1^{}+\lambda_2^{})
(\lambda_3^{}-\lambda_1^{})}} \; ,
\nonumber\\
O^{(0)}_{22} \hspace{-0.2cm} &=& \hspace{-0.2cm}
\sqrt{\frac{\lambda_2^{}
(\lambda_1^{}+\lambda_3^{})}{(\lambda_1^{}+\lambda_2^{})
(\lambda_2^{}+\lambda_3^{})}} \; ,
\nonumber\\
O^{(0)}_{23} \hspace{-0.2cm} &=& \hspace{-0.2cm}
\sqrt{\frac{(\lambda_2^{}-\lambda_1^{})
\lambda_3^{}}{(\lambda_3^{}-\lambda_1^{})
(\lambda_2^{}+\lambda_3^{})}} \; ,
\nonumber\\
O^{(0)}_{31} \hspace{-0.2cm} &=& \hspace{-0.2cm}
-\sqrt{\frac{\lambda_1^{} (\lambda_2^{} - \lambda_1^{})
(\lambda_1^{} + \lambda_3^{})}{(\lambda_1^{} + \lambda_2^{})
(\lambda_3^{} - \lambda_1^{}) (\lambda_1^{} - \lambda_2^{} +
\lambda_3^{})}} \; ,
\nonumber\\
O^{(0)}_{32} \hspace{-0.2cm} &=& \hspace{-0.2cm}
-\sqrt{\frac{\lambda_2^{} (\lambda_2^{} - \lambda_1^{})
(\lambda_3^{} - \lambda_2^{})}{(\lambda_1^{} + \lambda_2^{})
(\lambda_2^{} + \lambda_3^{}) (\lambda_1^{} - \lambda_2^{} +
\lambda_3^{})}} \; ,
\nonumber\\
O^{(0)}_{33} \hspace{-0.2cm} &=& \hspace{-0.2cm}
\sqrt{\frac{\lambda_3^{} (\lambda_1^{} + \lambda_3^{}) (\lambda_3^{}
- \lambda_2^{})}{(\lambda_3^{} - \lambda_1^{}) (\lambda_2^{} +
\lambda_3^{}) (\lambda_1^{} - \lambda_2^{} + \lambda_3^{})}} \; ,
\end{eqnarray}
where the superscript ``(0)" implies the Fritzsch (or $\epsilon^{}_a
= \epsilon^{}_b =0$) limit. Note that the above results hold for the
normal mass hierarchy (i.e., $\lambda^{}_1 < \lambda^{}_2 <
\lambda^{}_3$)
\footnote{This mass hierarchy is consistent with both the observed
mass spectra of charged fermions and the expected mass spectrum of
neutrinos. Although an inverse mass hierarchy is also possible for
three neutrinos, it cannot be consistent with the Fritzsch texture
$M^{(\rm F)}_\nu$ \cite{Xing02} or its non-Hermitian extension under
discussion.}.

Switching on the corrections of $\tilde{M}^{(\epsilon)}$ to
$\tilde{M}^{(\rm F)}$, one may calculate $O$ and $O^\prime$ appearing
in the bi-unitary transformation $O^\dagger \tilde{M}^{({\rm NNI})}
O^\prime = \widehat{M}^{}$ by following a perturbative way. We have
done such a perturbative calculation both to the first order of $\epsilon^{}_a$ and $\epsilon^{}_b$ and to the second order of
them, in order to examine whether the first-order analytical
approximations are good enough. Of course, the fermion mass hierarchies
should also be taken into account in our calculation. Given
$m^{}_e \ll m^{}_\mu \ll m^{}_\tau$, $m^{}_u \ll
m^{}_c \ll m^{}_t$ and $m^{}_d \ll m^{}_s \ll m^{}_b$, it is easy 
to simplify Eq. (7) by making reliable analytical approximations.
As only the normal hierarchy of three neutrino masses (i.e.,
$m^{}_1 < m^{}_2 <m^{}_3$) is allowed in this scenario, it is also
straightforward to simplify Eq. (7) in the neutrino sector.
This treatment might not be excellent if three neutrinos have a
relatively weak mass hierarchy, but it should be good enough for us
to reveal the salient features of non-symmetric corrections to the
Fritzsch textures. Our second-order analytical approximations in
diagonalizing $\tilde{M}^{(\rm NNI)} = \tilde{M}^{(\rm F)} + \tilde{M}^{(\epsilon)}$, which include the ${\cal O}(\epsilon^2_a)$
and ${\cal O}(\epsilon^2_b)$ corrections, support the above arguments
but they are too complicated to be presented here. To
the first order of $\epsilon^{}_a$ and $\epsilon^{}_b$, we simply
take $O = O^{(0)}\left({\bf 1} + X\right)$ and $O^\prime = O^{(0)}
\left({\bf 1} - X\right)$ with $X$ being anti-Hermitian (i.e.,
$X^\dagger = -X$) and proportional to the perturbation parameters
$\epsilon^{}_a$ and $\epsilon^{}_b$. In this case,
\begin{eqnarray}
\tilde{M}^{(\rm F)} \hspace{-0.2cm} &=& \hspace{-0.2cm}
O^{(0)} \ \widehat{M} \ {O^{(0)}}^\dagger \; ,
\nonumber \\
\tilde{M}^{(\epsilon)} \hspace{-0.2cm} &=& \hspace{-0.2cm}
O^{(0)} \left( X \widehat{M} + \widehat{M} X \right) {O^{(0)}}^\dagger \; .
\end{eqnarray}
Then we can determine the matrix elements of $X$ in terms of those
of $O^{(0)}$ and the perturbation parameters $\epsilon^{}_a$ and
$\epsilon^{}_b$. After an algebraic calculation, we obtain
\begin{eqnarray}
O_{11}^{} \hspace{-0.2cm} &=& \hspace{-0.2cm} O^{(0)}_{11}
\left[1-\frac{\lambda_1^{} \left[(\lambda_1^{}-\lambda_2^{})^2+(2
\lambda_1^{}-\lambda_2^{}) \lambda_3^{}+\lambda_3^2\right]
}{(\lambda_1^{}-\lambda_2^{}) (\lambda_1^{}+\lambda_3^{})
(\lambda_1^{}-\lambda_2^{}+\lambda_3^{})}{\epsilon^{}_a}
+\frac{\lambda_1^{}}
{\lambda_1^{}-\lambda_2^{}+\lambda_3^{}}{\epsilon^{}_b} \right] \; ,
\nonumber \\
O_{12}^{} \hspace{-0.2cm} &=& \hspace{-0.2cm} O^{(0)}_{12}
\left[1-\frac{\lambda_2^{}
\left[\lambda_1^2+(\lambda_2^{}-\lambda_3^{})^2-\lambda_1^{} (2
\lambda_2^{}-\lambda_3^{})\right]}{(\lambda_1^{}-\lambda_2^{})
(\lambda_2^{}-\lambda_3^{})
(\lambda_1^{}-\lambda_2^{}+\lambda_3^{})}{\epsilon^{}_a}-
\frac{\lambda_2^{}}{\lambda_1^{}-\lambda_2^{}+\lambda_3^{}}
{\epsilon^{}_b}\right] \; ,
\nonumber \\
O_{13}^{} \hspace{-0.2cm} &=& \hspace{-0.2cm} O^{(0)}_{13}
\left[1+\frac{\left[\lambda_1^2-\lambda_1^{} (\lambda_2^{}-2
\lambda_3^{})+(\lambda_2^{}-\lambda_3^{})^2\right] \lambda_3^{}
}{(\lambda_2^{}-\lambda_3^{}) (\lambda_1^{}+\lambda_3^{})
(\lambda_1^{}-\lambda_2^{}+\lambda_3^{})}{\epsilon^{}_a}+
\frac{\lambda_3^{}}{\lambda_1^{}-\lambda_2^{}+\lambda_3^{}}{\epsilon^{}_b}
\right] \; ,
\nonumber \\
O_{21}^{} \hspace{-0.2cm} &=& \hspace{-0.2cm} O^{(0)}_{21}
\left[1+\frac{\lambda_2^{} \lambda_3^{} (2
\lambda_1^{}-\lambda_2^{}+\lambda_3^{})}{(\lambda_1^{}-\lambda_2^{})
(\lambda_1^{}+\lambda_3^{}) (\lambda_1^{}-\lambda_2^{}+\lambda_3^{})}
{\epsilon^{}_a}-\frac{\lambda_1^{}}{\lambda_1^{}-\lambda_2^{}+\lambda_3^{}}
{\epsilon^{}_b} \right] \; ,
\nonumber \\
O_{22}^{} \hspace{-0.2cm} &=& \hspace{-0.2cm} O^{(0)}_{22}
\left[1+\frac{\lambda_1^{} \lambda_3^{} (\lambda_1^{}-2
\lambda_2^{}+\lambda_3^{}) }{(\lambda_1^{}-\lambda_2^{})
(\lambda_2^{}-\lambda_3^{}) (\lambda_1^{}-\lambda_2^{}+\lambda_3^{})}
{\epsilon^{}_a}+\frac{\lambda_2^{}}{\lambda_1^{}-\lambda_2^{}+\lambda_3^{}}
{\epsilon^{}_b}\right] \; ,
\nonumber \\
O_{23}^{} \hspace{-0.2cm} &=& \hspace{-0.2cm} O^{(0)}_{23}
\left[1-\frac{\lambda_1^{} \lambda_2^{} (\lambda_1^{}-\lambda_2^{}+2
\lambda_3^{})}{(\lambda_1^{}+\lambda_3^{})
(\lambda_2^{}-\lambda_3^{}) (\lambda_1^{}-\lambda_2^{}+\lambda_3^{})}
{\epsilon^{}_a}-\frac{\lambda_3^{}}{\lambda_1^{}-\lambda_2^{}+\lambda_3^{}}
{\epsilon^{}_b}\right] \; ,
\nonumber \\
O_{31}^{} \hspace{-0.2cm} &=& \hspace{-0.2cm} O^{(0)}_{31}
\left[1-\frac{\lambda_2^{} (\lambda_2^{}-\lambda_3^{}) \lambda_3^{}
}{(\lambda_1^{}-\lambda_2^{}) (\lambda_1^{}+\lambda_3^{})
(\lambda_1^{}-\lambda_2^{}+\lambda_3^{})}{\epsilon^{}_a}
+\frac{\lambda_2^{}-\lambda_3^{}}
{\lambda_1^{}-\lambda_2^{}+\lambda_3^{}}{\epsilon^{}_b} \right] \; ,
\nonumber\\
O_{32}^{} \hspace{-0.2cm} &=& \hspace{-0.2cm} O^{(0)}_{32}
\left[1+\frac{\lambda_1^{} \lambda_3^{} (\lambda_1^{}+\lambda_3^{})
}{(\lambda_1^{}-\lambda_2^{}) (\lambda_2^{}-\lambda_3^{})
(\lambda_1^{}-\lambda_2^{}+\lambda_3^{})}{\epsilon^{}_a}-
\frac{\lambda_1^{}+\lambda_3^{}}{\lambda_1^{}-\lambda_2^{}+\lambda_3^{}}
{\epsilon^{}_b}\right] \; ,
\nonumber \\
O_{33}^{} \hspace{-0.2cm} &=& \hspace{-0.2cm} O^{(0)}_{33}
\left[1-\frac{\lambda_1^{} (\lambda_1^{}-\lambda_2^{}) \lambda_2^{}
}{(\lambda_2^{}-\lambda_3^{}) (\lambda_1^{}+\lambda_3^{})
(\lambda_1^{}-\lambda_2^{}+\lambda_3^{})}{\epsilon^{}_a} -
\frac{\lambda_1^{} -\lambda_2^{} }{\lambda_1^{}
-\lambda_2^{}+\lambda_3^{}}{\epsilon^{}_b} \right] \; .
\end{eqnarray}
Given $\lambda_1^{} \ll \lambda^{}_3$ and $\lambda_2^{} \ll
\lambda_3^{}$, the above expressions of $O^{}_{ij}$ (for
$i,j=1,2,3$) may approximate to
\begin{eqnarray}
O^{}_{11} \hspace{-0.2cm} &\simeq& \hspace{-0.2cm}
\sqrt{\frac{\lambda_2^{}}{\lambda_1^{}+\lambda_2^{}}}\left(1+\frac{\lambda_1^{}
{}}{\lambda_2^{}-\lambda_1^{}}\epsilon^{}_a+\frac{\lambda_1^{}
}{\lambda_3^{}}{\epsilon^{}_b}\right) \; ,
\nonumber \\
O^{}_{12} \hspace{-0.2cm} &\simeq& \hspace{-0.2cm}
-\sqrt{\frac{\lambda_1^{}}{\lambda_1^{}+\lambda_2^{}}}\left(1-
\frac{\lambda_2^{}}{\lambda_2^{}-\lambda_1^{}}{\epsilon^{}_a}
-\frac{\lambda_2^{}}{\lambda_3^{}}{\epsilon^{}_b}\right) \; ,
\nonumber \\
O^{}_{13} \hspace{-0.2cm} &\simeq& \hspace{-0.2cm}
\sqrt{\frac{\lambda_1^{} \lambda_2^{}
(\lambda_2^{}-\lambda_1^{})}{\lambda_3^3}}\left(1-{\epsilon^{}_a}
+{\epsilon^{}_b}\right) \; ,
\nonumber \\
O^{}_{21} \hspace{-0.2cm} &\simeq& \hspace{-0.2cm}
\sqrt{\frac{\lambda_1^{} \lambda_3^{}}{(\lambda_1^{}+\lambda_2^{})(
\lambda_3^{}-\lambda_1^{}+\lambda_2^{})}}\left(1-\frac{\lambda_2^{}
}{\lambda_2^{}-\lambda_1^{}}{\epsilon^{}_a}-\frac{\lambda_1^{}
}{\lambda_3^{}}{\epsilon^{}_b}\right) \; , \nonumber \\ O^{}_{22}
\hspace{-0.2cm} &\simeq& \hspace{-0.2cm} \sqrt{\frac{\lambda_2^{}
\lambda_3^{}}{(\lambda_1^{}+\lambda_2^{})(
\lambda_3^{}-\lambda_1^{}+\lambda_2^{})}}\left(1+\frac{\lambda_1^{}{
}}{\lambda_2^{}-\lambda_1^{}}{\epsilon^{}_ a}+
\frac{\lambda_2^{}}{\lambda_3^{}}{\epsilon^{}_b}\right) \; ,
\nonumber\\
O^{}_{23} \hspace{-0.2cm} &\simeq& \hspace{-0.2cm}
\sqrt{\frac{\lambda_2^{}-\lambda_1^{}}{\lambda_3^{}-\lambda_1^{}
+\lambda_2^{}}}\left(1-{\epsilon^{}_b}\right) \; ,
\nonumber \\
O^{}_{31} \hspace{-0.2cm} &\simeq& \hspace{-0.2cm}
-\sqrt{\frac{\lambda_1^{}(\lambda_2^{}-\lambda_1^{})}{(\lambda_1^{}
+\lambda_2^{})(\lambda_3^{}-\lambda_1^{}-\lambda_2^{})}}
\left(1-\frac{\lambda_2^{}}{\lambda_2^{}-\lambda_1^{}}{\epsilon^{}_a}
-{\epsilon^{}_b}\right) \; ,
\nonumber \\
O^{}_{32} \hspace{-0.2cm} &\simeq& \hspace{-0.2cm}
-\sqrt{\frac{\lambda_2^{}(\lambda_2^{}-\lambda_1^{})}
{(\lambda_1^{}+\lambda_2^{})(\lambda_3^{}+\lambda_1^{}+\lambda_2^{})}}
\left(1+\frac{\lambda_1^{}}{\lambda_2^{}-\lambda_1^{}}
{\epsilon^{}_a}-{\epsilon^{}_b}\right) \; ,
\nonumber \\
O^{}_{33} \hspace{-0.2cm} &\simeq& \hspace{-0.2cm}
\sqrt{\frac{\lambda_3^{}}{\lambda_3^{}-\lambda_1^{}+\lambda_2^{}}}
\left(1+\frac{\lambda_2^{}-\lambda_1^{}
}{\lambda_3^{}}{\epsilon^{}_b}\right) \; .
\end{eqnarray}
It is obvious that the off-diagonal matrix elements of $O$ are
more sensitive to the corrections induced by the perturbation
parameters $\epsilon^{}_a$ and $\epsilon^{}_b$.

\section{A numerical illustration}

Now let us take a look at how sensitive the flavor mixing parameters
of leptons and quarks are to the perturbation parameters
$\epsilon^\alpha_a$ and $\epsilon^\alpha_b$ (for $\alpha ={\rm u},
{\rm d}; l, \nu$). We first discuss the CKM matrix $V$ and then
analyze the MNSP matrix $U$ in a numerical way.

\begin{center}
{\bf (A) The CKM matrix $V$}
\end{center}

Given six quark masses as the input parameters, the CKM matrix
$V=O^\dagger_{\rm u} P^{}_V O^{}_{\rm d}$ still contains six free
parameters: $\epsilon^{\rm u}_a$, $\epsilon^{\rm u}_b$,
$\epsilon^{\rm d}_a$, $\epsilon^{\rm d}_b$, $\phi^{}_1$ and
$\phi^{}_2$. Because the four perturbation parameters must be small,
we require $|\epsilon^{\rm u, d}_{a, b}| \lesssim 0.3$ as the
reasonable bounds in our numerical calculation. Then the
experimental data on four independent observable quantities of $V$,
typically chosen as $|V^{}_{us}|$, $|V^{}_{cb}|$, $|V^{}_{ub}|$ and
$\sin 2\beta$ with $\beta \equiv \arg\left[-(V^{}_{cd}
V^*_{cb})/(V^{}_{td} V^*_{tb})\right]$ being an inner angle of the
CKM unitarity triangle \cite{PDG}, will allow us to constrain the
parameter space of quark mass matrices $M^{(\rm NNI)}_{\rm u}$ and
$M^{(\rm NNI)}_{\rm d}$. Such a constraint will be useful for model
building.

To simplify the numerical calculation, we fix the values of quark
masses at the electroweak scale $\mu = M^{}_Z$ as follows: $m^{}_u =
2.0$ MeV, $m^{}_c = 0.557$ GeV, $m^{}_t = 168.3$ GeV; $m^{}_d = 2.7$
MeV, $m^{}_s = 47$ MeV and $m^{}_b = 2.92$ GeV \cite{XZZ,Branco}. In
addition, we adopt $|V^{}_{us}| = 0.2255 \pm 0.0019$, $|V^{}_{cb}| =
(41.2 \pm 1.1) \times 10^{-3}$, $|V^{}_{ub}| = (3.93 \pm 0.36)
\times 10^{-3}$ and $\sin 2\beta = 0.681 \pm 0.025$ \cite{PDG}. By
inputting the chosen values of six quark masses and allowing six
free parameters of $V$ to vary, one may then obtain the outputs of
$|V^{}_{us}|$, $|V^{}_{cb}|$, $|V^{}_{ub}|$ and $\sin 2\beta$ which
are required to lie in their respective ranges given above. This
treatment leads us to the parameter space of $\epsilon^{\rm u}_a$
versus $\epsilon^{\rm u}_b$, $\epsilon^{\rm d}_a$ versus
$\epsilon^{\rm d}_b$ and $\phi^{}_1$ versus $\phi^{}_2$, as shown in
Fig. 1. Some comments and discussions are in order
\footnote{Note that Branco {\it et al} have recently analyzed the
CKM matrix $V$ in such a way \cite{Branco}. Our more comprehensive
analysis not only confirms their results but also provides ourselves
with a meaningful calibration as we extend the same analysis to the
lepton sector. Our results for the MNSP matrix $U$ in section 3 (B)
are completely new.}.

(1) Although we have set the bounds $|\epsilon^{\rm u, d}_{a, b}|
\lesssim 0.3$, their allowed ranges are actually much smaller. In
particular, $\epsilon^{\rm u}_a \neq 0$ and $\epsilon^{\rm d}_b \neq
0$ hold, but $\epsilon^{\rm u}_b$ and $\epsilon^{\rm d}_a$ are
possible to vanish. This observation implies that both $M^{(\rm
NNI)}_{\rm u}$ and $M^{(\rm NNI)}_{\rm d}$ must be non-Hermitian. On
the other hand, one can see that $\epsilon^{\rm u}_a <0$ and
$\epsilon^{\rm d}_b >0$ hold. In comparison, $\epsilon^{\rm u}_b$ is
negative in most cases and $\epsilon^{\rm d}_a$ can be either
positive or negative. To reduce the number of free parameters from
four to two, one may either switch off $\epsilon^{\rm u}_b$ and
$\epsilon^{\rm d}_a$ or set $\epsilon^{\rm u}_a = \epsilon^{\rm
u}_b$ and $\epsilon^{\rm d}_a = \epsilon^{\rm d}_b$, or take
$\epsilon^{\rm u}_a = -\epsilon^{\rm d}_b$ and $\epsilon^{\rm u}_b =
-\epsilon^{\rm d}_a$, and so on. Such assumptions will strictly
constrain the textures of quark mass matrices and might be
suggestive for model building.

(2) It is impressive that two CP-violating phases $\phi^{}_1$ and
$\phi^{}_2$ are restricted to a quite narrow parameter space. In
particular, $|\phi^{}_1| \sim 90^\circ$ and $|\phi^{}_2| \sim
0^\circ$ imply that $\phi^{}_1$ dominates the strength of CP
violation in the CKM matrix $V$. This feature is similar to the one
showing up in some Hermitian modifications of the Fritzsch ansatz,
such as the four-zero textures of quark mass matrices \cite{4-zero}.

(3) In the quark sector we follow Ref. \cite{Branco} to define the
small parameter
\begin{eqnarray}
\epsilon \equiv \frac{1}{2} \sqrt{\left(\epsilon^{\rm u}_a \right)^2
+ \left(\epsilon^{\rm u}_b \right)^2
+ \left(\epsilon^{\rm d}_a \right)^2
+ \left(\epsilon^{\rm d}_b \right)^2} \;
\end{eqnarray}
to measure the overall non-Hermitian departure of $M^{(\rm
NNI)}_{\rm u}$ and $M^{(\rm NNI)}_{\rm d}$ from the Fritzsch
texture. The Jarlskog invariant of CP violation \cite{J}, defined as
$\cal J$ for the CKM matrix $V$, can be calculated via ${\cal J} =
{\rm Im}(V^{}_{us} V^{}_{cb} V^*_{ub} V^*_{cs})$. We illustrate the
numerical dependence of $\cal J$ on $\epsilon$ in Fig. 2, where
${\cal J} \sim 3.7 \times 10^{-5}$ for $\epsilon \sim 0.2$. The sign
of $\cal J$ is fixed by that of $\sin 2\beta$.

\begin{center}
{\bf (B) The MNSP matrix $U$}
\end{center}

Given three charged-lepton masses and two neutrino mass-squared
differences $\Delta m^2_{21} \equiv m^2_2 - m^2_1$ and $\Delta
m^2_{32} \equiv m^2_3 - m^2_2$, the MNSP matrix $U=O^\dagger_l
P^{}_U O^{}_\nu$ depends on seven free parameters: $\epsilon^l_a$,
$\epsilon^l_b$, $\epsilon^\nu_a$, $\epsilon^\nu_b$, $m^{}_1$,
$\varphi^{}_1$ and $\varphi^{}_2$. Again we require $|\epsilon^{l,
\nu}_{a, b}| \lesssim 0.3$ as the reasonable bounds. The present
neutrino oscillation data on three flavor mixing angles, denoted as
$\theta^{}_{12}$, $\theta^{}_{13}$ and $\theta^{}_{23}$ in the
standard parametrization of $U$ (i.e., $\tan\theta^{}_{12} =
|U^{}_{e 2}/U^{}_{e 1}|$, $\sin\theta^{}_{13} = |U^{}_{e3}|$ and
$\tan\theta^{}_{23} = |U^{}_{\mu 3}/U^{}_{\tau 3}|$) \cite{PDG},
will allow us to constrain the parameter space of lepton mass
matrices $M^{(\rm NNI)}_l$ and $M^{(\rm NNI)}_\nu$. For simplicity,
we fix the values of three charged-lepton masses at the electroweak
scale as follows: $m^{}_e = 0.48657$ MeV, $m^{}_\mu = 102.718$ MeV
and $m^{}_\tau = 1746.24$ MeV \cite{XZZ}. Moreover, we assume
$m^{}_1 = 0.0025$ eV and take $\Delta m^2_{21} = 8.0 \times 10^{-5}
~{\rm eV}^2$ and $\Delta m^2_{32} = 2.5 \times 10^{-3} ~ {\rm eV}^2$
together with $30^\circ < \theta^{}_{12} < 38^\circ$, $36^\circ <
\theta^{}_{23} < 54^\circ$ and $\theta^{}_{13} < 10^\circ$ in our
numerical calculation. Then the normal but weak neutrino mass
hierarchy is measured by two mass ratios $m^{}_1/m^{}_2 \simeq 0.27$
and $m^{}_2/m^{}_3 \simeq 0.18$. By inputting the chosen values of
charged-lepton and neutrinos masses and allowing the unknown
parameters of $U$ to vary, one may obtain the outputs of
$\theta^{}_{12}$, $\theta^{}_{13}$ and $\theta^{}_{23}$ which are
required to lie in their respective ranges given above. This
treatment leads us to the parameter space of $\epsilon^l_a$ versus
$\epsilon^l_b$, $\epsilon^\nu_a$ versus $\epsilon^\nu_b$ and
$\varphi^{}_1$ versus $\varphi^{}_2$, as shown in Fig. 3. Some
comments and discussions are in order.

(1) Because of $m^{}_e/m^{}_\mu \ll m^{}_1/m^{}_2$ and
$m^{}_\mu/m^{}_\tau \ll m^{}_2/m^{}_3$, the MNSP matrix $U$ is
expected to receive more contributions from the neutrino sector
rather than the charged-lepton sector. That is why the bounds on
$|\epsilon^l_{a, b}|$ are much looser than those on
$|\epsilon^\nu_{a, b}|$, as one can see from Fig. 3. We find that
$\epsilon^\nu_a$ and $\epsilon^\nu_b$ are negative in most cases. To
reduce the number of free parameters from four to two, one may
simply switch off $\epsilon^l_a$ and $\epsilon^l_b$. Although it is
also possible to set $\epsilon^\nu_a = \epsilon^\nu_b =0$, it will
be impossible to get $\theta^{}_{23} \simeq 45^\circ$ for the
atmospheric neutrino mixing angle \cite{Xing02}. More precise data
to be extracted from the upcoming neutrino oscillation experiments
may help us constrain the ranges of $\epsilon^l_{a, b}$ and
$\epsilon^\nu_{a, b}$ to a better degree of accuracy.

(2) Current neutrino oscillation experiments set no constraint on the
CP-violating phase $\varphi^{}_1$. But the other CP-violating phase
$\varphi^{}_2$ is well restricted to be around $180^\circ$, as shown
in Fig. 3. The reason is simply that a sufficiently large value of
$\theta^{}_{23}$ requires $\varphi^{}_2 \sim 180^\circ$. To see this
point more clearly, we write out
\begin{eqnarray}
\tan\theta_{23} = \left|\frac{U_{\mu 3}}{U_{\tau 3}}\right| \simeq
\left|\frac{h_l^{} \left(1-\epsilon_b^l\right)-
h_\nu^{}\left(1-\epsilon_b^\nu\right) e^{i \varphi^{}_2}}
{\left(1+h_l^2\epsilon_b^l+h_\nu^2\epsilon_b^\nu\right)+ h_l^{}
h_\nu^{} \left(1-\epsilon_b^l\right) \left(1-\epsilon_b^\nu\right)
e^{i \varphi^{}_2}}\right| \;
\end{eqnarray}
with the help of Eq. (10), where $h_l^{} \equiv
\sqrt{(m_\mu^{}-m_e^{})/m_\tau^{}} \simeq \sqrt{m_\mu^{}/m_\tau^{}}$
and $h_\nu^{} \equiv
\sqrt{(m_2^{}-m_1^{})/m_3^{}}~\hspace{-0.08cm}$. It becomes
transparent that $\varphi^{}_2 \sim 180^\circ$, together with
$\epsilon^l_b <0$ and $\epsilon^\nu_b <0$, may enhance the magnitude
of $\tan\theta^{}_{23}$ and make $\theta^{}_{23}$ closer to its
best-fit value (i.e., $\theta^{}_{23} \simeq 45^\circ$).

(3) Similar to the definition of $\epsilon$ in the quark sector,
a small parameter
\begin{eqnarray}
\epsilon^\prime \equiv \frac{1}{2} \sqrt{\left(\epsilon^l_a \right)^2
+ \left(\epsilon^l_b \right)^2
+ \left(\epsilon^\nu_a \right)^2
+ \left(\epsilon^\nu_b \right)^2} \;
\end{eqnarray}
can also be defined to measure the overall non-Hermitian departure
of $M^{(\rm NNI)}_l$ and $M^{(\rm NNI)}_\nu$ from the Fritzsch
texture. We illustrate the dependence of three flavor mixing angles
on $\epsilon^\prime$ in Fig. 4. The Jarlskog invariant of leptonic
CP violation, which can be calculated via ${\cal J}^\prime = {\rm
Im}(U^{}_{e2} U^{}_{\mu 3} U^*_{e 3} U^*_{\mu 2})$, is also shown in
Fig. 4. We see that it is possible to reach $|{\cal J}^\prime| \sim
1.5 \times 10^{-2}$. The CP-violating effects at this level should
be observable in the future long-baseline neutrino oscillation
experiments.

Finally, it makes sense to compare between the MNSP and CKM matrices
derived from the same NNI textures of lepton and quark mass matrices.
Given $|\epsilon^{\rm u,d}_{a,b}| \lesssim 0.3$ and
$|\epsilon^{l,\nu}_{a,b}| \lesssim 0.3$, it can be concluded that
the smallness of three quark mixing angles is primarily attributed to
the strong mass hierarchies of up- and down-type quarks, while the
largeness of solar and atmospheric neutrino mixing angles is mainly
ascribed to the relatively weak hierarchy of three neutrino
masses. Of course, the CP-violating phases play an important role
in either the lepton sector or the quark sector. Note that $V^{}_{ub}$
is smaller in magnitude than all the other elements of the CKM matrix
$V$, and $U^{}_{e3}$ is also the smallest element of the MNSP matrix
$U$. In other words, $\theta^{}_{13}$ is the smallest mixing angle in
both lepton and quark sectors. This interesting feature is a natural
consequence of the flavor textures of leptons and quarks together
with their corresponding mass hierarchies. In addition, the fact that
$|\epsilon^{\rm u,d}_{a,b}| \sim |\epsilon^{l,\nu}_{a,b}| \sim 0.2$
is favored by current experimental data should be quite suggestive of
a unified flavor model of leptons and quarks.

\section{Summary}

We have introduced non-Hermitian and nearest-neighbor-interacting
perturbations to the Fritzsch textures of lepton and quark mass
matrices such that both of them can fit current experimental data
very well. In particular, we find that it is possible to obtain
$\theta^{}_{23} \simeq 45^\circ$ for the atmospheric neutrino mixing
angle and predict $\theta^{}_{13} \simeq 3^\circ$ to $6^\circ$ for
the smallest neutrino mixing angle when the dimensionless
perturbation parameters in the lepton sector are at the $20\%$
level. We have shown that the same level of perturbations is
required in the quark sector, where the Jarlskog invariant of CP
violation is about $3.7 \times 10^{-5}$. In comparison, the strength
of leptonic CP violation is likely to reach about $1.5 \times
10^{-2}$ in neutrino oscillations.

As shown in Ref. \cite{Branco}, the NNI texture of quark mass
matrices can be derived from the introduction of an Abelian flavor
symmetry (e.g., the minimal realization of this idea requires a
$Z^{}_4$ flavor symmetry in the context of a two-Higgs doublet
model). We can follow the same procedure to obtain the NNI texture
of lepton mass matrices if massive neutrinos are the Dirac
particles. In the presence of a few heavy Majorana neutrinos, one
may first impose the aforementioned flavor symmetry on the Yukawa
interaction of neutrinos to get the NNI texture for the Dirac
neutrino mass matrix and then achieve the Majorana mass matrix for
three light neutrinos via the seesaw mechanism. There are therefore
a number of possibilities of model building, but the numerical
results must be very different from what we have presented in this
work. One may explore such possibilities once more accurate
experimental data on neutrino masses and lepton flavor mixing angles
are available in the (near) future, and in particular when the simple
scenario discussed in this paper is phenomenologically discarded or
becomes less favored.

Let us reiterate that a parallel study of lepton and quark mass
matrices, such as the approximate Fritzsch textures under
discussion, is useful to reveal the similarities and differences
between the lepton and quark sectors. It should also be helpful for
building a unified flavor model of leptons and quarks with the help
of proper flavor symmetries.

\vspace{0.4cm}

This work was supported in part by the National Natural Science
Foundation of China under grant No. 10875131 and in part by the
Ministry of Science and Technology of China under grant No.
2009CB825207.

\newpage

\begin{figure}
\center
\begin{overpic}[width=12cm,height=22cm]{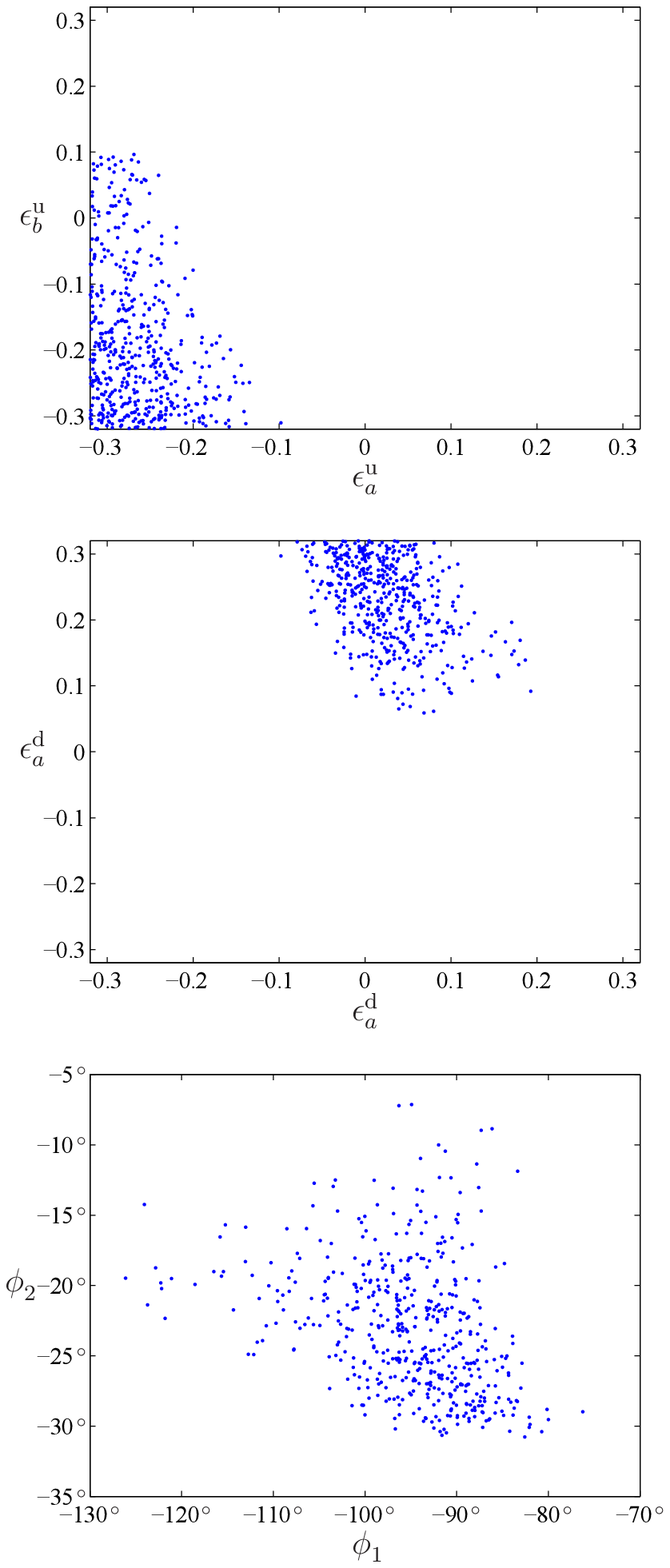}
\end{overpic}
\caption{An illustration of the parameter space of $\epsilon^{\rm
u}_a$ versus $\epsilon^{\rm u}_b$, $\epsilon^{\rm d}_a$ versus
$\epsilon^{\rm d}_b$ and $\phi^{}_1$ versus $\phi^{}_2$ constrained
by current data in the quark sector.}
\end{figure}

\begin{figure}
\center
\begin{overpic}[width=14cm]{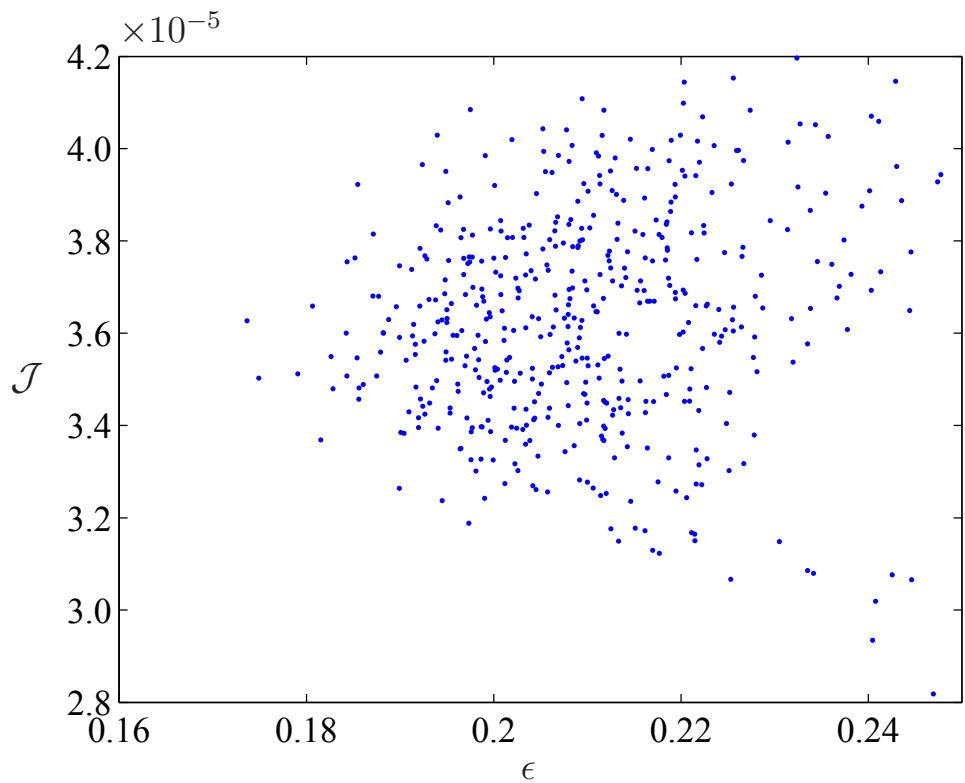}
\end{overpic}
\caption{An illustration of the dependence of the Jarlskog invariant
of CP violation $\mathcal{J}$ on the overall perturbation parameter
$\epsilon$ in the quark sector.}
\end{figure}

\begin{figure}
\center
\begin{overpic}[width=12cm,height=22cm]{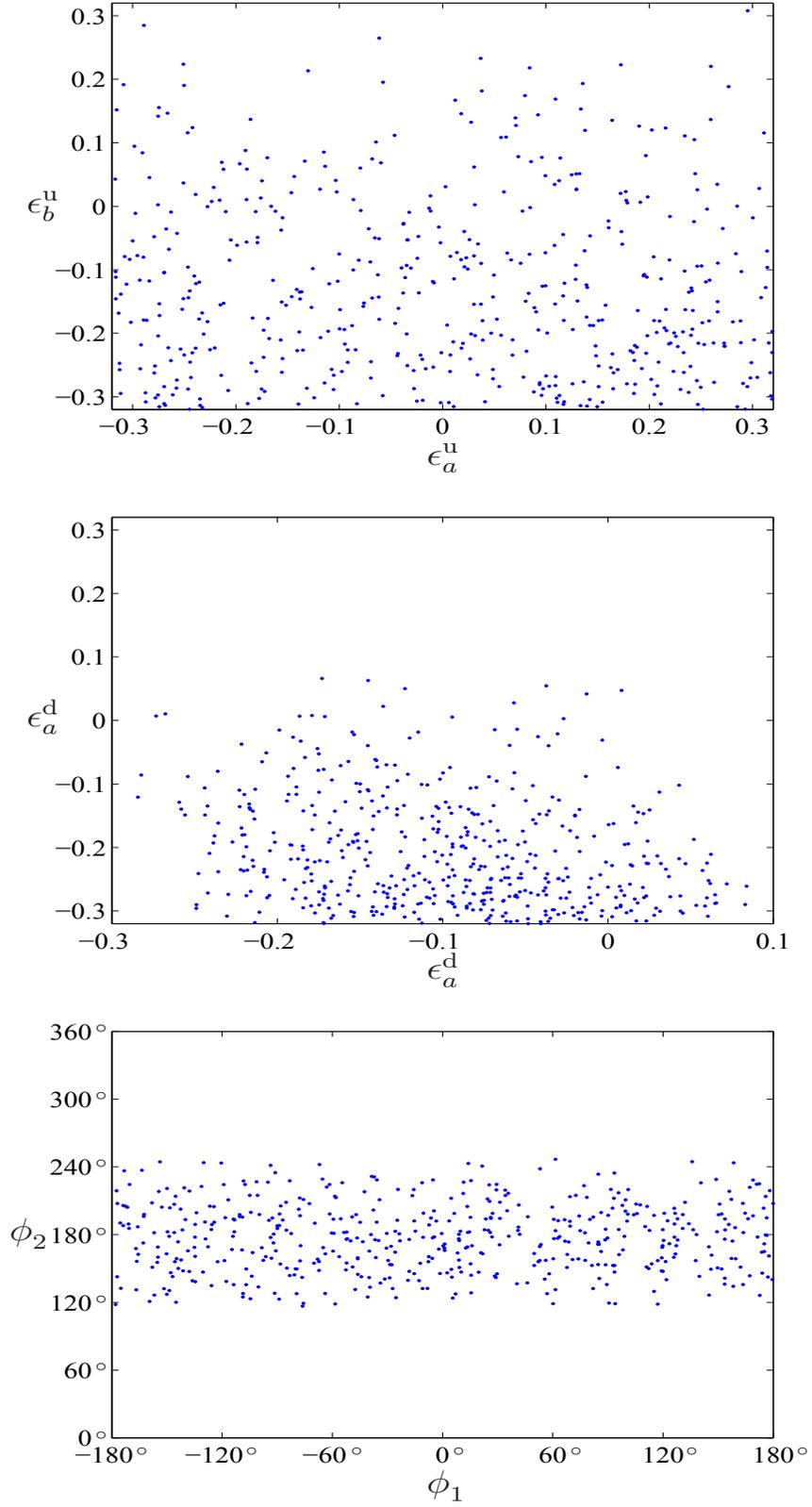}
\end{overpic}
\caption{An illustration of the parameter space of $\epsilon^l_a$
versus $\epsilon^l_b$, $\epsilon^\nu_a$ versus $\epsilon^\nu_b$ and
$\varphi^{}_1$ versus $\varphi^{}_2$ in the lepton sector with the
input $m^{}_1 = 0.0025$ eV.}
\end{figure}

\begin{figure}
\center
\begin{overpic}[width=14cm]{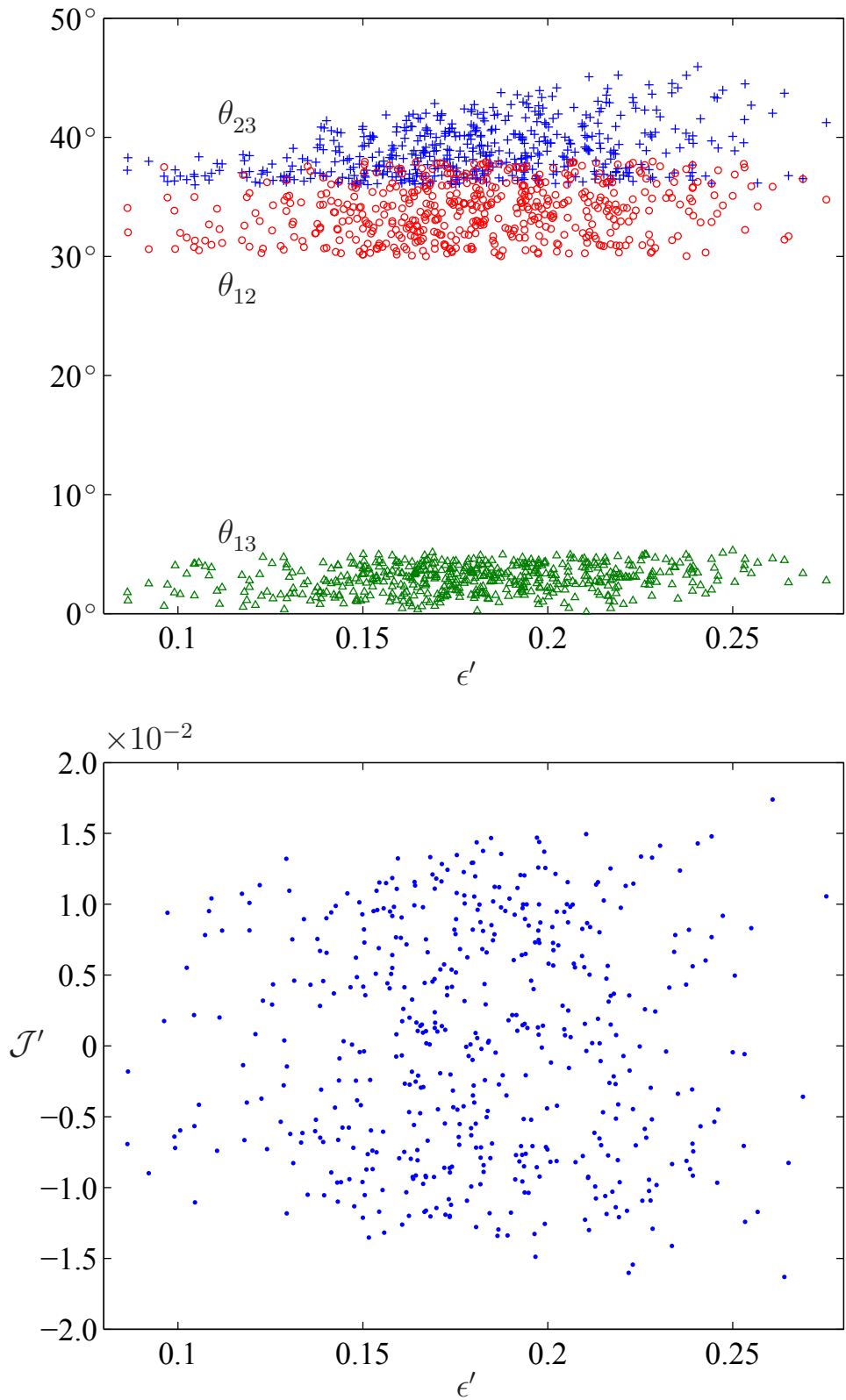}
\end{overpic}
\caption{An illustration of the dependence of three flavor
mixing angles and the Jarlskog invariant of CP violation ${\cal
J}^\prime$ on the overall perturbation parameter $\epsilon^\prime$
in the lepton sector.}
\end{figure}

\end{document}